\documentclass[5p,,preprint,12pt,twocolumn]{elsarticle}
\makeatletter\if@twocolumn\PassOptionsToPackage{switch}{lineno}\else\fi\makeatother

\usepackage{tabulary,xcolor}
\usepackage{amsfonts,amsmath,amssymb}
\usepackage[T1]{fontenc}
\makeatletter
\let\save@ps@pprintTitle\ps@pprintTitle
\def\ps@pprintTitle{\save@ps@pprintTitle\gdef\@oddfoot{\footnotesize\itshape \null\hfill\today}}
\def\hlinewd#1{%
  \noalign{\ifnum0=`}\fi\hrule \@height #1%
  \futurelet\reserved@a\@xhline}

\AtBeginDocument{\ifNAT@numbers \biboptions{sort&compress}\fi}
\makeatother

\usepackage{amsmath}
\usepackage{graphicx}
\usepackage[colorinlistoftodos]{todonotes}
\usepackage[colorlinks=true, allcolors=black]{hyperref}
\usepackage{float}
\usepackage{lipsum}
\usepackage{subcaption}
\usepackage{relsize}
\usepackage{dblfloatfix} 
\usepackage{kantlipsum}
\usepackage{nccmath}

\usepackage{ifluatex}
\ifluatex
\usepackage{fontspec}
\defaultfontfeatures{Ligatures=TeX}
\usepackage[]{unicode-math}
\unimathsetup{math-style=TeX}
\else 
\usepackage[utf8]{inputenc}
\fi 
\ifluatex\else\usepackage{stmaryrd}\fi

\usepackage{url,multirow,morefloats,floatflt,cancel,tfrupee}
\makeatletter

\AtBeginDocument{\@ifpackageloaded{textcomp}{}{\usepackage{textcomp}}}
\makeatother
\usepackage{colortbl}
\usepackage{xcolor}
\usepackage{pifont}
\usepackage[nointegrals]{wasysym}
\makeatletter

\def\mcWidth#1{\csname TY@F#1\endcsname+\tabcolsep}

\def\cAlignHack{\rightskip\@flushglue\leftskip\@flushglue\parindent\z@\parfillskip\z@skip}
\def\rAlignHack{\rightskip\z@skip\leftskip\@flushglue \parindent\z@\parfillskip\z@skip}

\@ifundefined{etal}{}{}

\usepackage{ifxetex}
\ifxetex\else\if@twocolumn\@ifpackageloaded{stfloats}{}{\usepackage{dblfloatfix}}\fi\fi

\AtBeginDocument{
\expandafter\ifx\csname eqalign\endcsname\relax
\def\eqalign#1{\null\vcenter{\def\\{\cr}\openup\jot\m@th
  \ialign{\strut$\displaystyle{##}$\hfil&$\displaystyle{{}##}$\hfil
      \crcr#1\crcr}}\,}
\fi
}

\AtBeginDocument{%
  \@ifpackageloaded{endfloat}%
   {\renewcommand\efloat@iwrite[1]{\immediate\expandafter\protected@write\csname efloat@post#1\endcsname{}}}{\newif\ifefloat@tables}%
}%

\def\BreakURLText#1{\@tfor\brk@tempa:=#1\do{\brk@tempa\hskip0pt}}
\let\lt=<
\let\gt=>
\def\processVert{\ifmmode|\else\textbar\fi}

\@ifundefined{subparagraph}{
\def\subparagraph{\@startsection{paragraph}{5}{2\parindent}{0ex plus 0.1ex minus 0.1ex}%
{0ex}{\normalfont\small\itshape}}%
}{}

\newcommand\role[1]{\unskip}
\newcommand\aucollab[1]{\unskip}
  
\@ifundefined{tsGraphicsScaleX}{\gdef\tsGraphicsScaleX{1}}{}
\@ifundefined{tsGraphicsScaleY}{\gdef\tsGraphicsScaleY{.9}}{}
\def\checkGraphicsWidth{\ifdim\Gin@nat@width>\linewidth
	\tsGraphicsScaleX\linewidth\else\Gin@nat@width\fi}

\def\checkGraphicsHeight{\ifdim\Gin@nat@height>.9\textheight
	\tsGraphicsScaleY\textheight\else\Gin@nat@height\fi}

\def\fixFloatSize#1{}
\let\ts@includegraphics\includegraphics

\def\inlinegraphic[#1]#2{{\edef\@tempa{#1}\edef\baseline@shift{\ifx\@tempa\@empty0\else#1\fi}\edef\tempZ{\the\numexpr(\numexpr(\baseline@shift*\f@size/100))}\protect\raisebox{\tempZ pt}{\ts@includegraphics{#2}}}}

\AtBeginDocument{\def\includegraphics{\@ifnextchar[{\ts@includegraphics}{\ts@includegraphics[width=\checkGraphicsWidth,height=\checkGraphicsHeight,keepaspectratio]}}}

\DeclareMathAlphabet{\mathpzc}{OT1}{pzc}{m}{it}

\def\URL#1#2{\@ifundefined{href}{#2}{\href{#1}{#2}}}

\def\UrlOrds{\do\*\do\-\do\~\do\'\do\"\do\-}%
\g@addto@macro{\UrlBreaks}{\UrlOrds}

\edef\fntEncoding{\f@encoding}

\makeatother

\newif\ifmultipleabstract\multipleabstractfalse%
\begin{document}

\begin{frontmatter}
	
\title{A mathematical model to describe the alpha dose rate\\ from a UO$_2$ surface}
    
\author[a99b9588a3165]{Angus Siberry}
\ead{as14659@bristol.ac.uk}
\author[a13f753534016]{David Hambley}
\author[af593676aeb90]{Anna Adamska}
\author[a99b9588a3165]{Ross Springell}
    
\address[a99b9588a3165]{
    University of Bristol\unskip, HH Wills Physics Laboratory\unskip, Bristol\unskip, BS8 1TL\unskip, UK. \\
    }
  	
\address[a13f753534016]{ 
    National Nuclear Laboratory Ltd\unskip, Central Laboratory\unskip, Sellafield\unskip, CA20 1PG\unskip, Cumbria\unskip, UK. \\
    }
  	
\address[af593676aeb90]{
 Sellafield Ltd\unskip, Sir Christopher Harding House\unskip, Whitehaven\unskip, CA28 7XY\unskip, UK. \\
    }

\begin{abstract}
A model to determine the dose rate of a planar alpha-emitting surface, has been developed. The approach presented is a computationally efficient mathematical model using stopping range data from the Stopping Ranges of Ions in Matter (SRIM) software. The alpha dose rates as a function of distance from irradiated UO$_2$ spent fuel surfaces were produced for benchmarking with previous modelling attempts. This method is able to replicate a Monte Carlo (MCNPX) study of an irradiated UO$_2$ fuel surface within 0.6 \% of the resulting total dose rate and displays a similar dose profile.

\end{abstract}
\end{frontmatter}

\section*{Keywords}
Alpha Radiation, Dosimetry, UO$_2$, Radiolysis, SRIM, Nuclear Materials
\section{Introduction}

The role of nuclear power in its potential to combat climate change is well-established \cite{innovation2020research}. Despite this, due to high cost and concerns over safety, its future as a major energy resource is uncertain \cite{verbruggen2008renewable}. A drawback of nuclear power is the complex waste forms that it produces, and the potential decommissioning challenges associated with radioactive materials \cite{seier2014environmental}. To develop a comprehensive plan of how to deal with this waste there must be an understanding of what could happen when moving, treating and storing such waste. In order to do this safely, predictive tools are required to highlight the potential risks and how to mitigate them. \\
\indent An important component of nuclear power production is the management of spent fuel. Whether the fuel is to be reprocessed or placed in a geological disposal facility, the maintenance and assessment of fuel integrity during storage is crucially important. Upon exposure to water, dissolution of the fuel matrix and a release of highly radioactive fission products can occur \cite{repository_2000,poinssot2005spent,shoesmith2007used}. In many storage practices this exposure is possible. In the case of a geological repository, it is even expected due to the large time scales associated with the fuel being in one location. A detailed understanding of these degradation mechanisms and the conditions that drive them could improve the effectiveness of any control measures, influence facility design and ultimately, reduce the cost. Developing modelling tools to predict the rate of radioactive dose and dose profile through the fuel-water interface is a critical part of a wider effort to provide accurate predictions of fuel dissolution rates in the event of a containment breach.\\
\indent Spent nuclear fuel consists of predominantly, UO$_2$ ($\approx$ 95 \%); the remaining material is comprised of fission products and other actinides \cite{bobrowski2017application}. The reactivity of UO$_2$ in water is so low it is almost considered inert \cite{SHOESMITH1996287}, however, if oxidised the uranium valence state converts from U(IV) to the much more soluble U(VI); hence, in the presence of oxidising species, UO$_2$ will corrode more rapidly leading to a faster release of the radioactive isotopes held within the fuel matrix \cite{shoesmith2007used,bright2019comparing,SPRINGELL}.\\
\indent G-value is the yield of a particular species resulting from ionisation \cite{bobrowski2017application, buxton2008overview}. It can be used for relating instantaneous yields, normally of radical species, or equilibrium yield of molecular species. The G-value used commonly in disposal environments is the yield
of molecular species, used to convert the energy lost by ionising radiation in water, to the number of molecules of a given species produced. This is the method whereby dosimetry results can be used to determine dissolution kinetics in chemical reaction and diffusion models \cite{shoesmith2007used,Jonsson_steady,Jonssonh2o2,Jonsson_time}.\\ \indent Alpha particles can generate energetic species which are able to react with each other and their surrounding environment \cite{draganic1971radiation,elliot2009reaction}. This process can produce oxidising conditions near the solid-water interface, which, due to the short penetration depth of alpha particles in water, varies rapidly with distance from the surface.  In many senses beta and gamma radiation is more ubiquitous than alpha across the fuel cycle because they are more penetrating. However, the more rapid decay of beta and gamma emitting nuclides results in alpha radiation being dominant at the fuel interface when considering the timescales relevant to disposal environments (>1000 yrs) \cite{Nazhen_Shoesmith_Review}. These considerations indicate the importance of developing an accurate model of alpha radiation across the fuel-water interface.\\ \indent Most radiolysis models utilise the linear energy transfer (LET) curve for calculating the dose received by a medium per decay. This is because the rate at which a corpuscle is stopped is equivalent to the rate the energy is transferred to the medium; hence, a linear energy transfer between the two. The functional form of the LET is described by the following relation.
\begin{equation}
    LET = - \frac{dE}{dx}
\end{equation}
where E is the energy lost by the ion and x is the length over which it is lost. High LET radiation refers to slower heavier ions and low refers to fast moving electrons. To model radiolysis from high LET radiation the chemical events occur so frequently along its path we assume a uniform cylindrical region, known as the penumbra \cite{elliot2009reaction}. In order to obtain the LET function of a single alpha particle, the stopping powers of each material it is traversing through is required. The stopping power of a material can be derived from Bethe-Bloch theory.

\subsection{Stopping power}
\begin{figure*}[!t]
    \centering
    \includegraphics[width=0.98\textwidth]{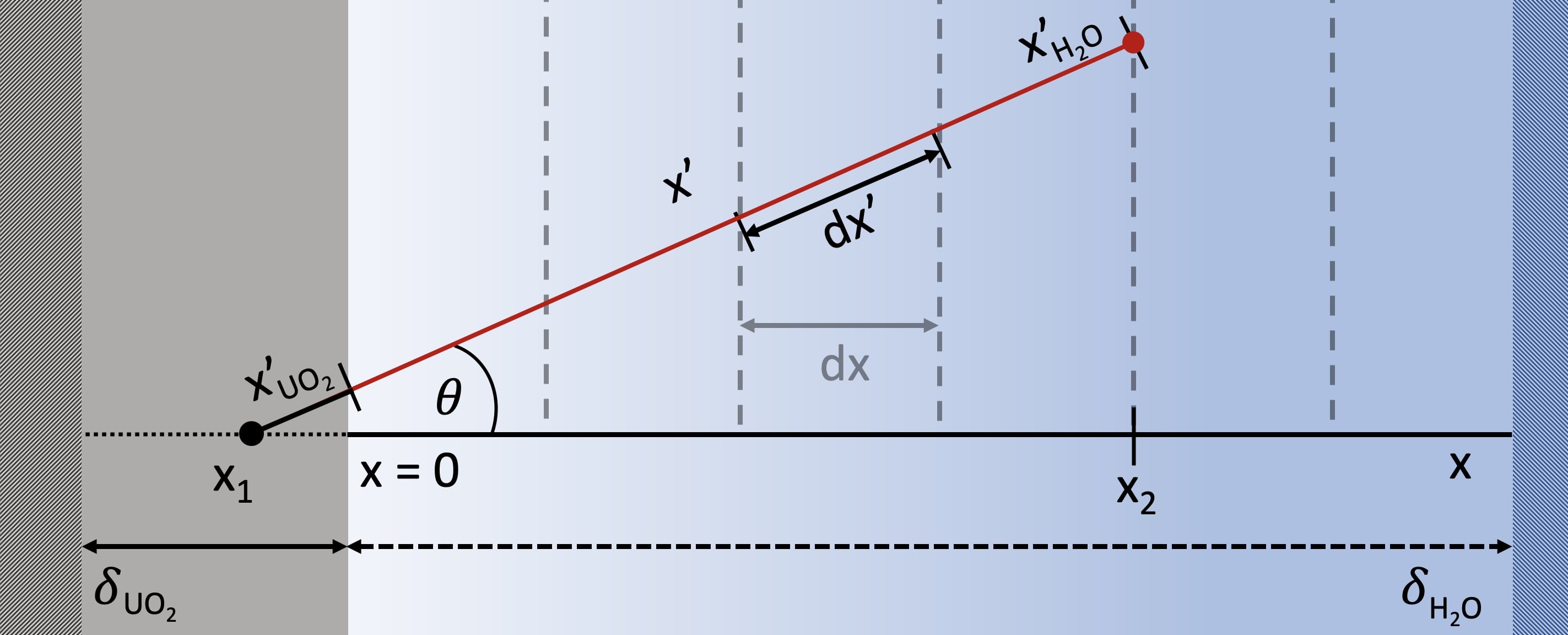}
    \caption{An illustration of the geometry used in the planar surface model. The x$_1$ position indicates a randomly generated position on the dotted line, while the x$_2$ position indicates the perpendicular range of the alpha particle. The solid black-red line denoted x'$_{UO_2}$  and  x'$_{H_2O}$ indicates the distances travelled along the axis x', a randomly generated path at an angle $\theta$ from the x axis, in UO$_2$ and H$_2$O, respectively. dx and dx' represent infinitesimal distances between successive layers in H$_2$O.}
    \label{fig:Linear}
    \vspace{2mm}
 \hrule
\end{figure*}
Bethe-Bloch theory describes the average energy lost by a charged particle due to Coulomb interactions between the particle and the electrons of atoms within the medium \cite{grimes2017approximate}. At the basis of all stopping power models, lies the Bethe-Bloch equation \cite{bloch1933bremsvermogen}. The equation, with additional corrections, does well to predict the stopping ranges of high velocity ions through a variety of media \cite{lindhard1963range}.  The Stopping Ranges of Ions in Matter (SRIM), a program created by Ziegler and Biersack, contains a comprehensive database of experimental values to use alongside a corrected Bethe-Bloch model \cite{ziegler2008srim}. SRIM generates the stopping corrections required from compounds containing common elements. This process is known as the core and bond (CAB) approach. It uses the interaction between the traversing ion in the atomic centres and adding the stopping from the materials bonding electrons \cite{ziegler2008srim}. The accuracy of this SRIM software had been tested through many compounds \cite{H_ion_Sore,He_andersen,H_ion_Au_ISHI,montanari2017iaea} and found to predict the stopping of H and He ions within 2 \% at the Bragg peak \cite{ziegler2008srim}. The LET curve can be extracted from the ionisation output.

\subsection{Modelling approaches}
The modelling approaches used for determining alpha dose rates can be split into two categories: analytical derivations; utilising stopping power ratios, tables and geometries \cite{sunder1998calculation,nielsen2006geometrical,poulesquen2006spherical,hansson2020alpha}, or Monte Carlo methods utilising nuclear Monte Carlo simulators such as GEANT4 or MCNP \cite{kumazaki2007determination,miller2006MCNP,mougnaud2015glass,tribet2017spent}. The Monte Carlo simulators are often computationally demanding whereas the analytical approach often oversimplifies stopping power and refrains from treating the energy distribution of particles separately. The model described in this study utilises the SRIM software, alongside geometrical considerations, in an attempt to produce a fast and accurate method for determining dose rates from planar alpha emitting surfaces, particularly UO$_2$. This study also highlights issues in dimensional analysis when simulating this geometry and contradicts a theoretical analysis by Hansson \textit{et al.} \cite{hansson2020alpha}. The following model was built in Python with the use of the math, random and Numpy libraries.

\section{Methods and calculations}
\subsection{Geometry}
\begin{figure*}[!b]
  \centering
  \includegraphics[width=\linewidth]{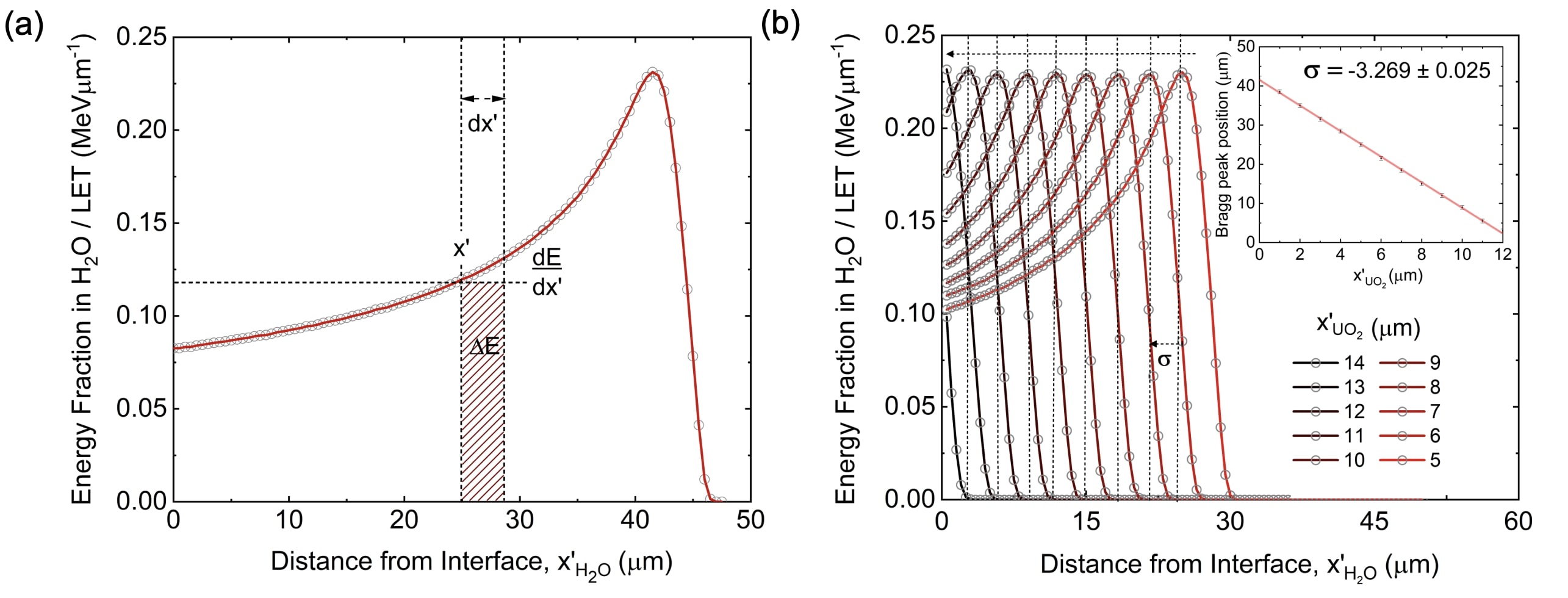}
\caption{(a) Illustration of energy interval deposited at decay path displacement $x'$ with a width $dx'$ that is deposited at distance $x$ from the interface. (b) Illustration of Bragg curve through $x'_{H_2O}$ with peak shift denoted, $\sigma$. The quantity $\sigma$ highlighted as a function of distance travelled within UO$_2$, denoted $x'_{UO_2}$. Both illustrations use an alpha particle with a decay energy of 5.8 MeV.}
\label{fig:bragg_curves}
 \vspace{2mm}
 \hrule
\end{figure*}

The most commonly built spent fuel dosimetry model is that of a planar surface of UO$_2$. The maximum thickness of UO$_2$ considered, is bound by the furthest distance an alpha particle can travel through the medium, $\delta_{UO_2}$, with a given decay energy. The water layer is bound similarly, but that of a maximum distance, $\delta_{H_2O}$. The setup of this model is illustrated in Figure \ref{fig:Linear}. The illustration shows the dependence on the path length $x'$ in each medium at a given decay depth $x_1$. To calculate the dose as a function of distance from the surface, a summation of all decay paths, $x'$, and the associated LET fraction deposited within each interval $dx$ (see Figure \ref{fig:bragg_curves}) needs to be made. The dose at a distance $x$ from the surface for a given decay is given as
\begin{equation}
    D(x)   =  \frac{e}{A\rho_{H_2O}}\int^{x + \frac{dx}{2}}_{x- \frac{dx}{2}}{\frac{dE}{dx'}dx'}
\end{equation}
where $\rho$ is the density of the medium, A the area of surface in question and the factor, e, is used to determine the dose in joules. To convert the dose into a dose rate you need the flux of ions. The flux of ions is dependent on the activity and geometry of the source. Assuming a radioactive surface emitting radiation in 1D, the surface flux can be defined as  
\begin{equation}
F  = a \rho_{src} V_{src}P_{\alpha},
\end{equation}
where $a$ is the specific activity in Bqg$^{-1}$, $\rho_{src}$ and $V_{src}$ are the density and volume of the source material considered, respectively. The volume of the source material is bound by the maximum depth an alpha can originate from in the fuel and still contribute to a dose at or beyond the surface, this is denoted $\delta_{UO_2}$ in Figure \ref{fig:Linear}. Considering now the direction of each decay, the average probability of an alpha escaping within V$_{src}$ must be considered. This is because at least 50 \% of all decays will go back in the direction of the bulk. This quantity can be defined by the summation of escape probabilities at each depth, $x_1$. This property will be denoted $P\alpha$ and can be calculated by the following derivation.
\begin{figure}[!b]
    \centering
    \includegraphics[width=\linewidth]{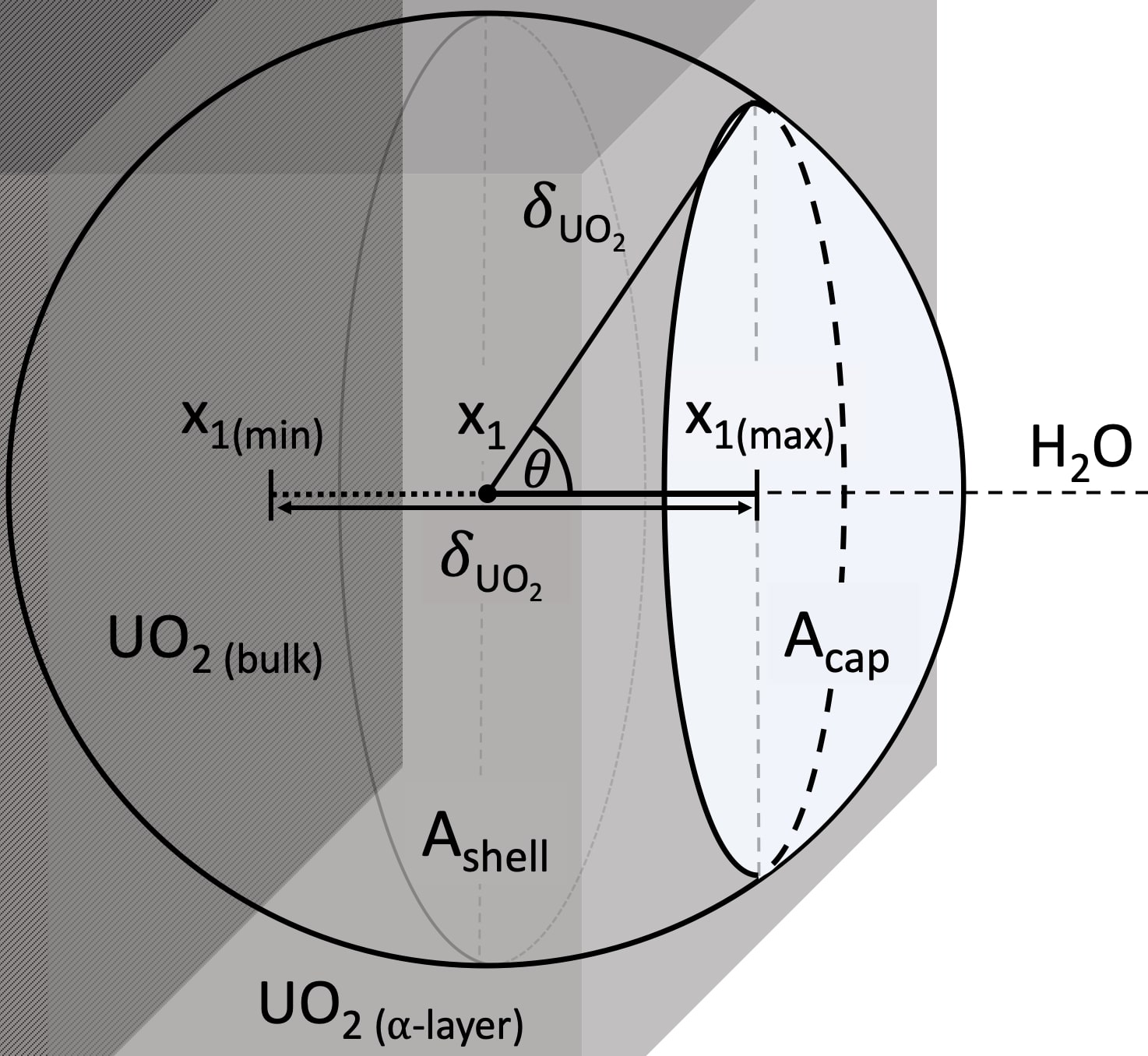}
    \caption{An illustration showing the parameters required to calculate the average probability of escape. A sphere of radius $\delta_{UO_2}$ is bound by the centre point, x$_1$. The position x$_1$ is bound between the depth of $\delta_{UO_2}$ denoted x$_{1(min)}$ and the UO$_2$ surface denoted x$_{1(max)}$. The angle, $\theta$, represents the maximum angle from the x axis by which a decay trajectory can escape the UO$_2$ surface at the starting depth x$_1$. The quantities A$_{shell}$ and A$_{cap}$ indicate the surface area of the sphere and cap that is exposed beyond the UO$_2$ surface.}
    \label{fig:decay_sphere}
\end{figure}

\subsection{Probability of alpha escape}
Considering the limits of probable escape, we can deduce that at the surface there is a 50 \% chance of the alpha particle leaving the material. At the maximum depth defined by $\delta_{UO_2}$ we deduce a 0 \% chance beyond this. Now assuming that the path is straight for every decay and that the $\theta$ dependence is truly stochastic, we can assume isotropic decays of length $\delta_{UO_2}$ in all directions, forming a spherical `shell' of probable decay positions. As shown in the derivation by Nielsen \textit{et al}. \cite{nielsen2006geometrical}, the probability of the decay breaching the surface is given by the ratio of that spherical shell surface area which lies beyond the surface, to the total shell surface area. An illustration of this is shown in Figure \ref{fig:decay_sphere}. 

For a cap of radius $\delta_{UO_2}$, and height $(\delta_{UO_2} - x_1)$ where $x_1$ is the distance from the sphere to the intersecting plane, the surface area is given by
\begin{equation}
A_{cap} = 2\pi \delta_{UO_2}(\delta_{UO_2} - x_1).  
\end{equation}
If the probability of alpha escape is the ratio of the cap to the decay shell, it is given by the following relation
\begin{equation}
    P\alpha(x_1) = \frac{A_{cap}}{A_{shell}} = \frac{(\delta_{UO_2} - x_1)}{2\delta_{UO_2}}.
\end{equation}
this is a linear function with an average value of 0.25.

\subsection{Dose rate}
The dose rate, \.{D}, over a defined interval at $x$ can be estimated using the dose of an alpha particle  travelling though $x$ multiplied by the total flux from the surface. Combining (2) and (3) the dose rate becomes
\begin{equation}
    \dot{D} = \frac{e a \rho_{UO_2}\delta_{UO_2}P\alpha  }{\rho_{H_2O}}\int^{x + \frac{dx}{2}}_{x- \frac {dx}{2}}{\frac{dE}{dx'}dx'}.
\end{equation}

Each $\alpha$ particle is considered to have a different trajectory (Figure \ref{fig:Linear}). In order to model energy deposition as a function of perpendicular distance from the surface, each particle must be treated separately. To model dose 
in the unit of Gys$^{-1}$  and setting P$\alpha = 0.25$, the number of decays within 1 s would be
\begin{equation}
    n = \frac{a\delta_{UO_2}\rho_{UO_2}}{4}
\end{equation}
  where n is the total number of particles emitted from the UO$_2$ surface per second. Therefore, for a dose rate (in Gy$s^{-1}$) at x distance from the UO$_2$ surface and a water layer width of dx, the resulting equation becomes
\begin{equation}
     \dot{D} = \frac{e}{\rho_{H_2O}} \mathlarger{\mathlarger{\sum}} ^{n = \mathlarger{\frac{a\delta_{UO_2}\rho_{UO_2}}{4}}}_{\mathlarger{1}}\mathlarger{\int}^{x + \frac{dx}{2}}_{x- \frac {dx}{2}}{\frac{dE_n}{dx_n'}dx_n'}
\end{equation}
where,
\begin{equation}
    x = x'\cos{\theta}.
\end{equation}
The term $\frac{dE_n}{dx_n'}$ should considered as the LET of the $n$th particle emitted from the UO$_2$ surface at a distance $x'$ through its trajectory (see Figure \ref{fig:bragg_curves}a).

\subsection{SRIM}
When considering the LET of an alpha particle the distance travelled within each medium is of particular importance. In the case of received dose in water the limiting factor is the energy of the alpha particle as it crosses the fuel-water interface. This will be dependent on the characteristic decay energy and distance travelled in the UO$_2$ medium. An assumption has been made that all alpha particles have a characteristic initial decay energy for simplification within the model. The energy deposited per alpha will in this case be bound by the distance travelled within UO$_2$. The effect this has on the LET function within water is shown in Figure \ref{fig:bragg_curves}b, illustrated by the peak shift, $\sigma$. Since the Bragg curve maintains much of its functional form, and the peak shifts in a linear fashion with distance travelled in UO$_2$, a `base-function' of a Bragg curve unimpeded by UO$_2$ can be used to then approximate all other Bragg curves in this model. This is only valid as long as the peak shift $\sigma$ is well-understood. 

\subsection{Peak shift $\sigma$}
 Figure \ref{fig:bragg_curves}b shows the Bragg peak shift, $\sigma$, of an alpha particle with a decay energy of 5.8 MeV, that occurs with increased path distance within the UO$_2$ medium ($x'_{UO_2}$) before crossing the interface into H$_2$O. This shift shows a strong negative linear correlation with $x'_{UO_2}$ as previously shown by Poulesquen and J\'{e}gou \cite{poulesquen2007linear}. A line of best fit of the Bragg peak position over 12 values of $x'_{UO_2}$ (see Figure \ref{fig:bragg_curves}b), $\sigma$ is approximated by the gradient at -3.269$\pm$0.025 with an R$^2$ value of 0.999.

Understanding this shift is key in reducing the complexity and computation time of this model. Instead of simulating the LET interaction at a time, as a function of depth and angle of emission for each particle (through both the UO$_2$ and H$_2$O medium), all that is required is the functional form of the Bragg curve unimpinged by UO$_2$ ($x'_{UO_2}=0$), and the total distance travelled in UO$_2$. The Bragg curve of an alpha particle through water only will be referred to as the `base-function' throughout.

Once the peak shift and base-function of the Bragg curve for an alpha particle of a given energy are known, the model can be built using matrices and linear algebra in a 2D Cartesian geometry.
 
\section{Results}
\begin{figure}[!t]
    \centering
    \includegraphics[width=\linewidth]{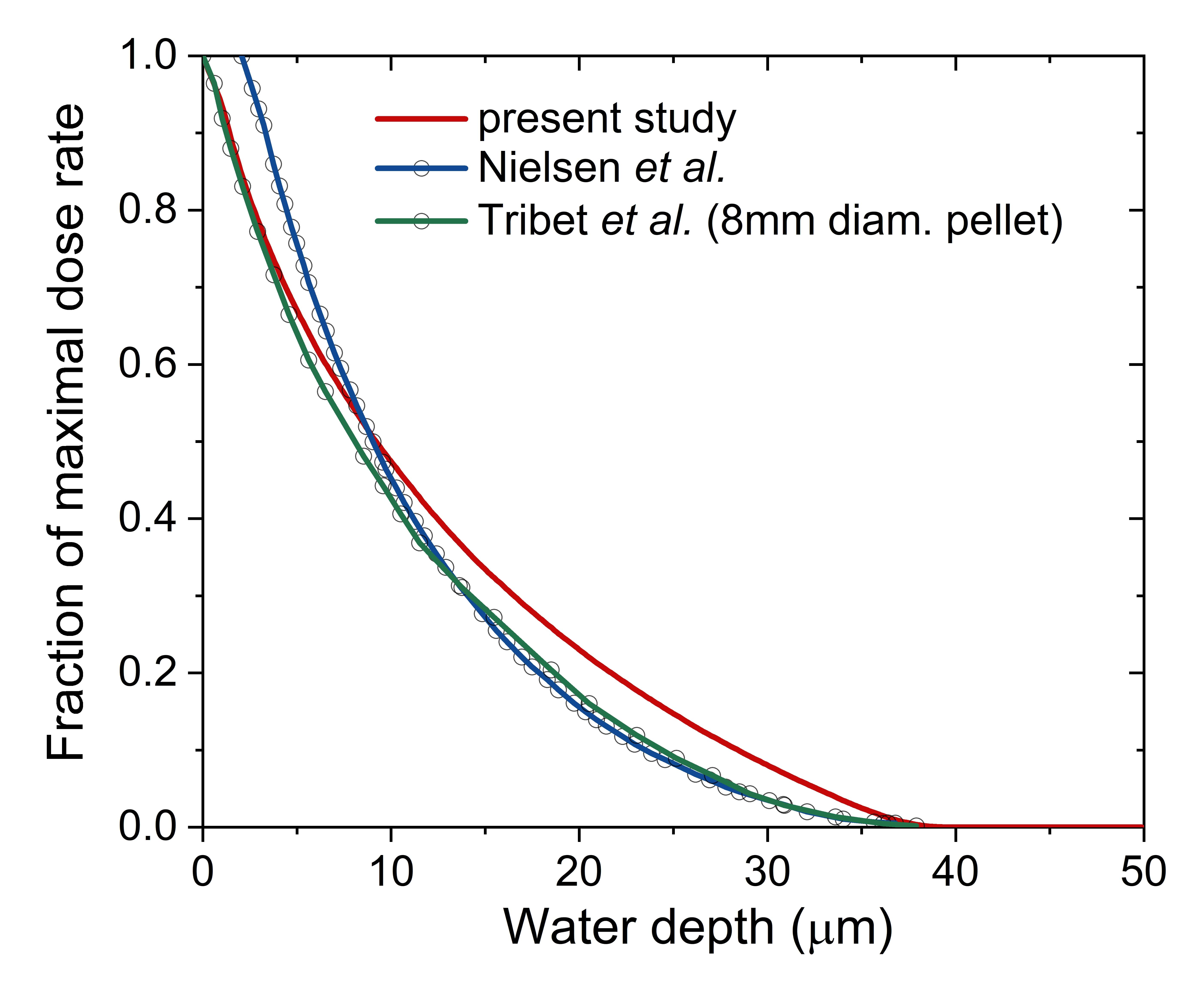}
    \caption{Graph showing a comparison of the results from the planar geometry model in the present study against previous literature. Due to the variety of activities chosen in the literature the fraction of maximal dose rate is used.}
    \label{fig:linear_comp}
\end{figure}

\begin{figure}[!b]
    \centering
    \includegraphics[width=\linewidth]{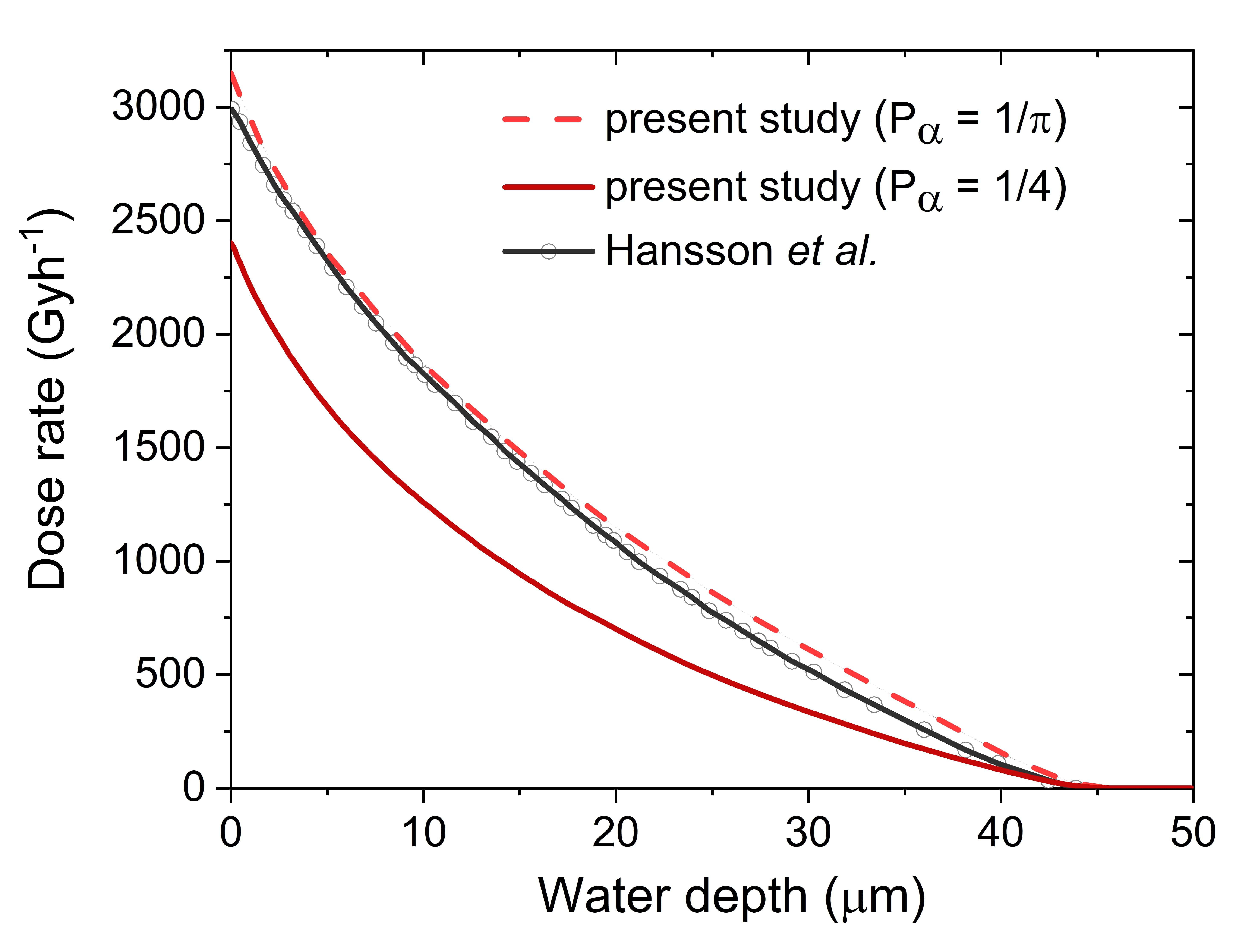}
    \caption{Graph showing the Hansson model result (solid black line) and the present study for decays of 5.8 MeV and an activity of 5.6$\times 10^8$ Bqg$^{-1}$. The dashed line shows results from the present study using the Hansson assumption of P$_{\alpha} = 1/\pi$ and the solid red line indicating the present study with the corrected P$_{\alpha} = 1/4$.}
    \label{fig:Hansson}
\end{figure}

Figure \ref{fig:linear_comp} shows the resulting dose rate in the form of a decay curve that aligns well with previous theoretical studies \cite{dzaugis2015quantitative,hansson2020alpha,nielsen2006geometrical,tribet2017spent}. The fraction of the peak dose rate is used to make the comparison activity-independent. The simulation number used to run each decay and scale to the appropriate dose used is 100,000. Results remain relatively similar with increased computational cost after this amount as the model converges to within a 0.5 \% fluctuation after 50,000.\\
\begin{figure}[!t]
    \centering
    \includegraphics[width=\linewidth]{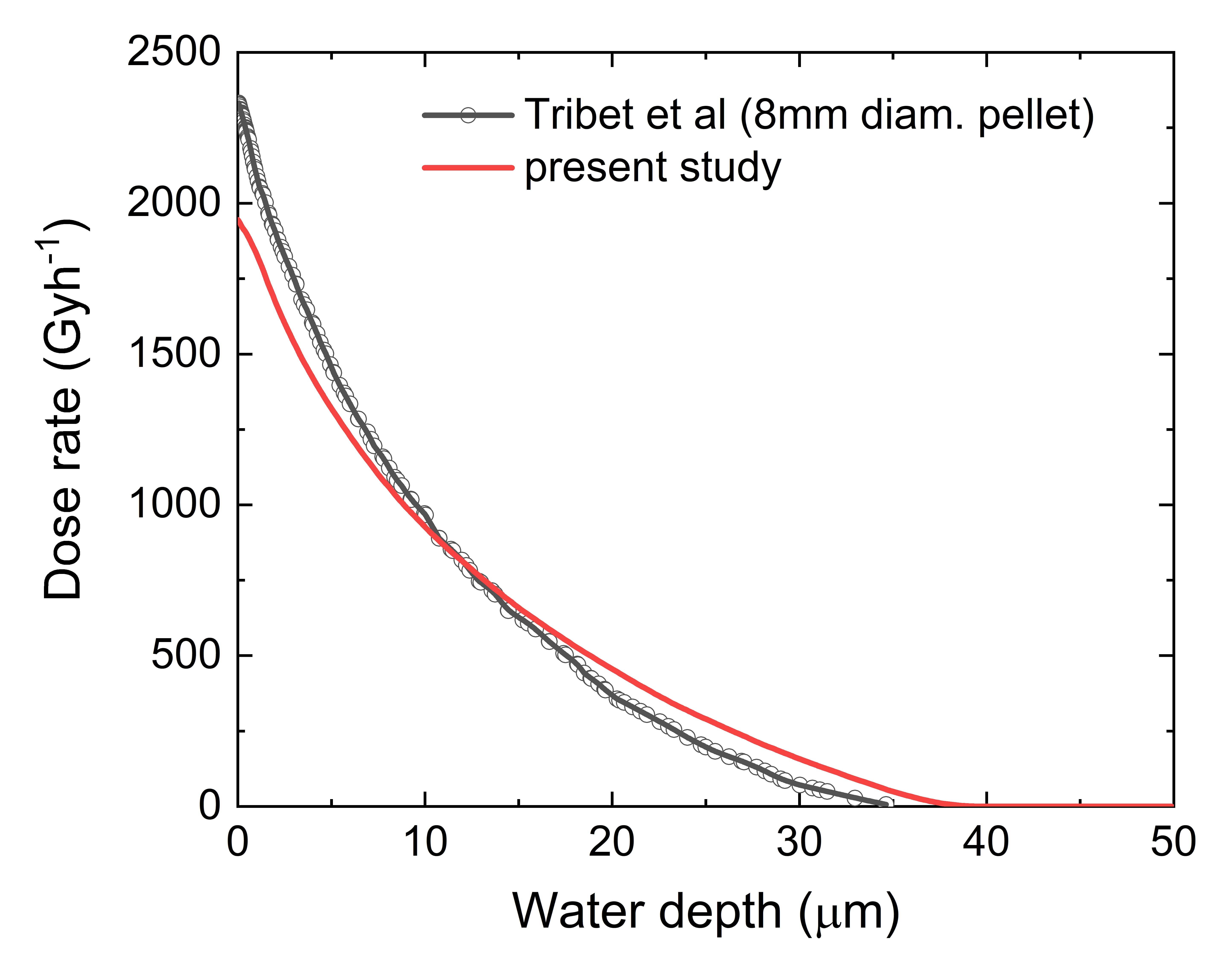}
    \caption{Graph showing the Tribet model result and the present study for decays of 5.3 MeV and an activity of 4.73$\times 10^8$ Bqg$^{-1}$.}
    \label{fig:Tribet}
\end{figure}
\indent To determine how well the model performs with regard to the magnitude of dose, we can see a direct comparison in Figures \ref{fig:Tribet} and \ref{fig:Hansson}. The total dose of the model built by Tribet \textit{et al.} was calculated to be 23740 Gyh$^{-1}$, in comparison with this study a value of 23880 Gyh$^{-1}$ is in good agreement. A comparison of average values over 30 $\mu$m from previous studies are shown in Table 1. As expected, the values presented in this study and the values by Tribet \textit{et al.} are in good agreement. In comparison to the model by Hansson \textit{et al.} \cite{hansson2020alpha}, the Bethe-Bloch LS calculation presented by Cachoir \textit{et al.} \cite{Cachior}, and the estimation made in the SFS report \cite{poinssot2005spent}, our results suggests a significant overestimation in average dose rates from these studies.

\subsection{Hansson model}
\begin{table*}
\centering
\renewcommand{\arraystretch}{1.2}
\begin{tabular}{c|c|c|c|c|c}
 Decay Energy  & Tribet \textit{et al.} & SFS Report & Cachoir \textit{et al.} & Hansson \textit{et al.} & present study \\
 (Mev) & (Gyh$^{-1}$)& (Gyh$^{-1}$)& (Gyh$^{-1}$)& (Gyh$^{-1}$)& (Gyh$^{-1}$)\\\hline
 5.3 & 791 & - & -  & - & 731 \\
 5.8 & - &  1760  & 1680 & 1628 & 1136 \\\hline

\end{tabular}

\caption{A literature comparison of the average dose rate values in water over 30$\mu$m from the fuel-water interface \cite{tribet2017spent,Cachior,hansson2020alpha}. The studies shown that use decay energies 5.3 and 5.8 Mev use an activity of 4.73$\times 10^8$ Bqg$^{-1}$ and 5.6$\times 10^8$ Bqg$^{-1}$, respectively.}

\end{table*}

When analysing the literature there seems to be a mistaken assumption about the probability of escape in the Hansson paper \cite{hansson2020alpha}. The paper points out that Hosoe \textit{et al.} \cite{hosoe1984stopping} correctly stated that 25 \% of all particles escape a planar surface. The article goes on to mention that a paper by Garisto \textit{et al.} \cite{garisto1989energy} `corrected' the assumption that the energy of the alpha particle was linear and instead, a function of emission angle. It goes on to say that if the energy-dependence of the alpha particle travelling through the medium was a function of emission angle, then the escape probability should also follow the same angular-dependence. Garisto concluded that the “The self-shielding by the surface layer of the fuel reduces the total number and energy of $\alpha$ particles emerging from the fuel surface by a factor of four and seven, respectively” , making the distinction between energy-dependence and particle escaping \cite{garisto1989energy}.
 This was overlooked and instead Hansson theorised that because the energy-dependence follows a cosine function the average value of $\cos{\theta}$ between $0 < \theta <\pi / 2$ ($4/\pi$) multiplied by the previously assumed 0.25 will correctly give an escape probability of $1/\pi$ (0.318). This was backed up by the fact that their model computationally produced a similar escape probability (0.315). Unfortunately, this result is due to a dimensional analysis error. The model describes a 2D geometrical setup. To most simply calculate the escape probability, one takes all the possible escape positions limited by $\delta$ and divide by all possible end positions (see A.1. for full derivation). This is the ratio of a circle's arc (where $r = \delta$), with its circumference in 2D. The probability gives the following relation:
\begin{equation}
    P = \frac{cos^{-1}{(|x| - \delta})}{\pi}
\end{equation}
This is a non-linear function with an average value between $-\delta < x < 0$ of $1/\pi$. If the model was set up in 3D the calculated value would have been as predicted, P$_\alpha = 0.25$, as shown in the derivation in Section 2.2.

\section{Discussion}
The dose rate profile was calculated for alpha particles with initial decay energies of 5.8 MeV and 5.3 MeV, emitted from an infinitely planar surface, using a geometrical model in Python with the aid of fitted data from SRIM. In this model the resulting distributions resemble decay curves with the  maximal dose rate at the fuel-water interface. The resulting decay curve for initial energies of 5.3 MeV were compared with previous attempts by Nielsen \textit{et al.} and Tribet \textit{et al.} \cite{nielsen2006geometrical,tribet2017spent} (Figure \ref{fig:linear_comp}).\\
\indent The Nielsen model derives the alpha dosimetry rate and curve from the energy-dependence of the particle range formula created by Jansson and Jonsson \cite{JannsonJonsson}. This technique has been shown to underestimate the range compared to the SRIM software by Hansson \textit{et al.} \cite{hansson2020alpha}. A potentially key feature of the dose rate curve is the interface dose rate, as this represents the local rate of radiolysis, and hence the generation of highly reactive radicals that could affect the corrosion of the surface material \cite{Nazhen_Shoesmith_Review}. As this model does not calculate a dose rate closer than 3 $\mu$m from the surface it could lead to an underestimation of the corrosion rate. However, it has been shown that using the average values of dose rate within the alpha-irradiated volume one can simulate corrosion kinetics in good agreement with experimental data \cite{Jonssonh2o2,Jonsson_time,Jonsson_steady}. The role of dose rate shape on the rate of dissolution has not been properly investigated and may be of greater importance in more complex geometries.\\
\indent In a report from the SFS project (an EU framework 6 project) an estimation was made of the interface and total dose rate of alpha particles emitted from a UO$_2$ surface. They used an initial energy 5.8 MeV and activity 5.6$\times 10^{8}$ Bqg$^{-1}$ \cite{poinssot2005spent}, the same parameters used in the comparison with Hansson (Figure \ref{fig:Hansson}). The interface dose rate estimate was 3120 Gyh$^{-1}$, which was compared to a model using the Bethe-Bloch LS equation by Cachoir \textit{et al.} \cite{Cachior}. The Cachoir model greatly overestimated the dose rate, while the report alluded to an interface dose rate that Hansson closely predicts. Nevertheless, this is still an overestimation of the interface dose rate.\\
\indent The estimation uses consecutive layers of material each emitting a dose of 1022 Gyh$^{-1}$ in the direction of the interface. It then multiplies the total fuel layers contributing to a dose beyond the surface then divides by the total water layers receiving the dose. This oversimplification fails to consider attenuation of neighbouring fuel layers reducing the dose rate with each consecutive layer going deeper into the fuel. Hence the interface estimation is again an overestimation, further alluding to the validity of this model and the error in P$_{\alpha}$ made in the Hansson report, this is supported by the difference in average dose values shown in Table 1. The 27 \%  difference in total dose received due to the P$_{\alpha}$ correction could considerably the effect of result of dissolution or radiolysis modelling built on such a result.\\ 
\indent The dose rate curve by Tribet was simulated using the Monte Carlo N Particle (MCNPX) transport code that evaluates all particle interactions at set layer-widths and approximates between layers \cite{tribet2017spent}. In a comparison to the results in this study (Figure \ref{fig:linear_comp}), they produce logical values with a slight deviation from this model beyond 10 $\mu$m. Despite the disparity in decay shape, the total dose is in good agreement with Tribet with a larger range (Figure \ref{fig:Tribet}), and a lower interface dose rate, due to underlying differences in the stopping ranges used for 5.3 MeV. A slight difference in the two models is the use of UO$_2$ density as 10.8 gcm$^{-3}$ ($\mathrm{^{238}UO_2}$) by Tribet, instead of the 10.97 gcm$^{-3}$ used in this study. Comparing the density effect on the stopping range calculated by SRIM equates to an increase of 1.5 \% in the stopping range (12.35 $\mu$m to 12.54 $\mu$m) due to a reduction in density. The approach presented in this study is therefore in good agreement with the MCNPX model, while improving computational efficiency. The similarity in average values within an alpha-irradiated volume of thickness 30 $\mu$m shown in Table 1 also support the argument for implementation of this model over the MCNPX approach.\\  \indent There are significant computational challenges when combining dose rate calculations with a chemical reaction and diffusion model. To overcome this, dose rate calculations are often significantly simplified using values determined analytically \cite{Nazhen_Shoesmith_Review,Jonsson_steady,Jonsson_time,Cachior}. If the approach presented in this study was combined with a chemical reaction and diffusion model one could, in comparison to previous analytical approaches, better predict the dissolution rate of spent fuel and hydrogen release as a function of fuel age and repository condition.

\section{Conclusions}
This study presents a mathematical model, producing dose rate curves with computational ease and built on a simple geometrical approach with the use of fitted SRIM data. The model performs in good agreement with Tribet but differs to the Hansson model due to their overestimation in the average probability of alpha escape. Alpha dose rate models are of particular importance for the study of spent fuel-water interface behaviour. To better understand the importance of dose-rate curve shape and the role it plays on the dissolution of spent fuel, a deeper understanding of the dissolution mechanisms of the fuel-water interface is required. \\ 

\section*{Acknowledgements}
This work was supported by The Engineering and Physical Sciences Research Council (EPSRC) and the Transformative Science and Engineering for Nuclear Decommissioning (TRANSCEND) consortium.

\appendix 

\section{Appendix}\label{appendix-title-5bbcd27cbcac}
\subsection{The role of dimension on P$_\alpha$}   
\begin{figure}[!t]
    \centering
    \includegraphics[width=\linewidth]{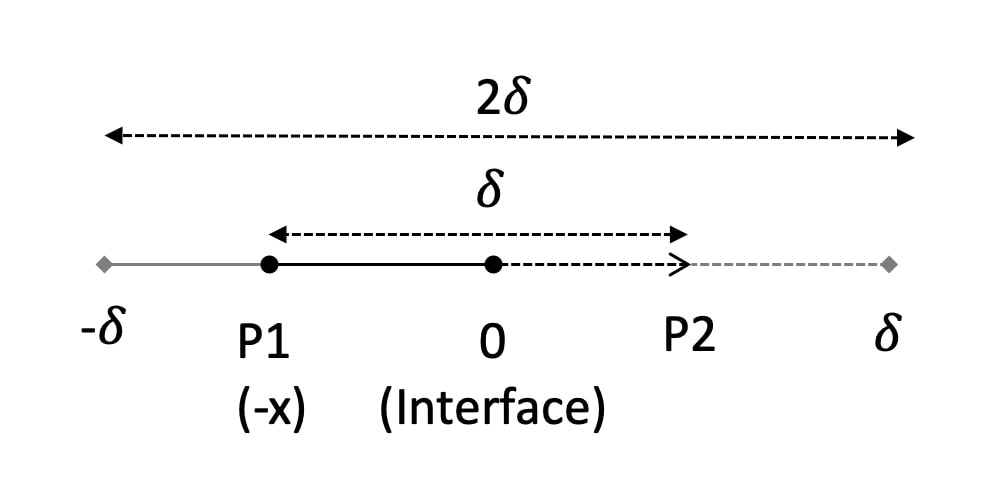}
    \caption{An illustration of the parameters used for the probability of alpha escape function for 1-dimension.}
    \label{fig:1D}
\end{figure}

For the 1-Dimensional system shown in Figure  \ref{fig:1D} , the probability of P1 to P2 crossing the interface if it can go forwards and backwards by the distance, $\delta$, is the length of the path past interface divided by the full range of the particle, $2\delta$. This gives the following probability function for a starting position $-\delta < x < 0$,
\begin{equation}
P = \frac{1}{2} - \frac{|x|}{2\delta}
\end{equation}
it can be clearly seen that this is a linear function with respect to x, where the average probability is found at the midpoint of $x$, where $x = -\frac{\delta}{2}$. Giving an average probability of escape equal to $1/4$.

\begin{figure}[!t]
    \centering
    \includegraphics[width=\linewidth]{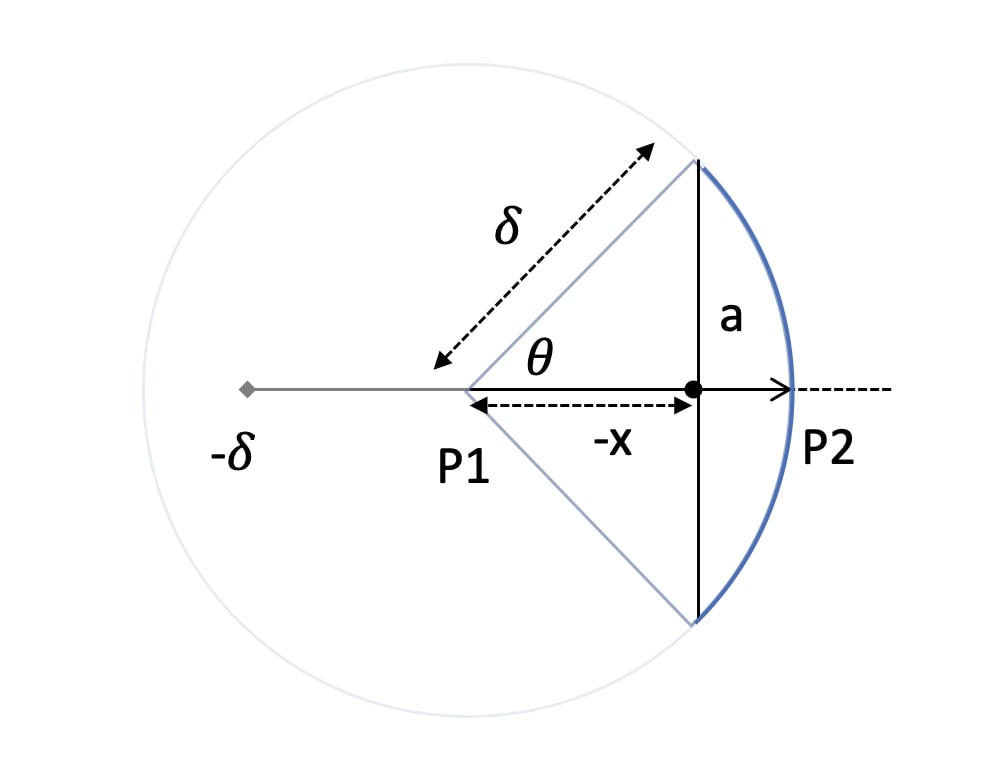}
    \caption{An illustration of the parameters used for the probability of alpha escape function for 2-dimensions.}
    \label{fig:2D}
\end{figure}
\begin{figure}[!b]
    \centering
    \includegraphics[width=0.9\linewidth]{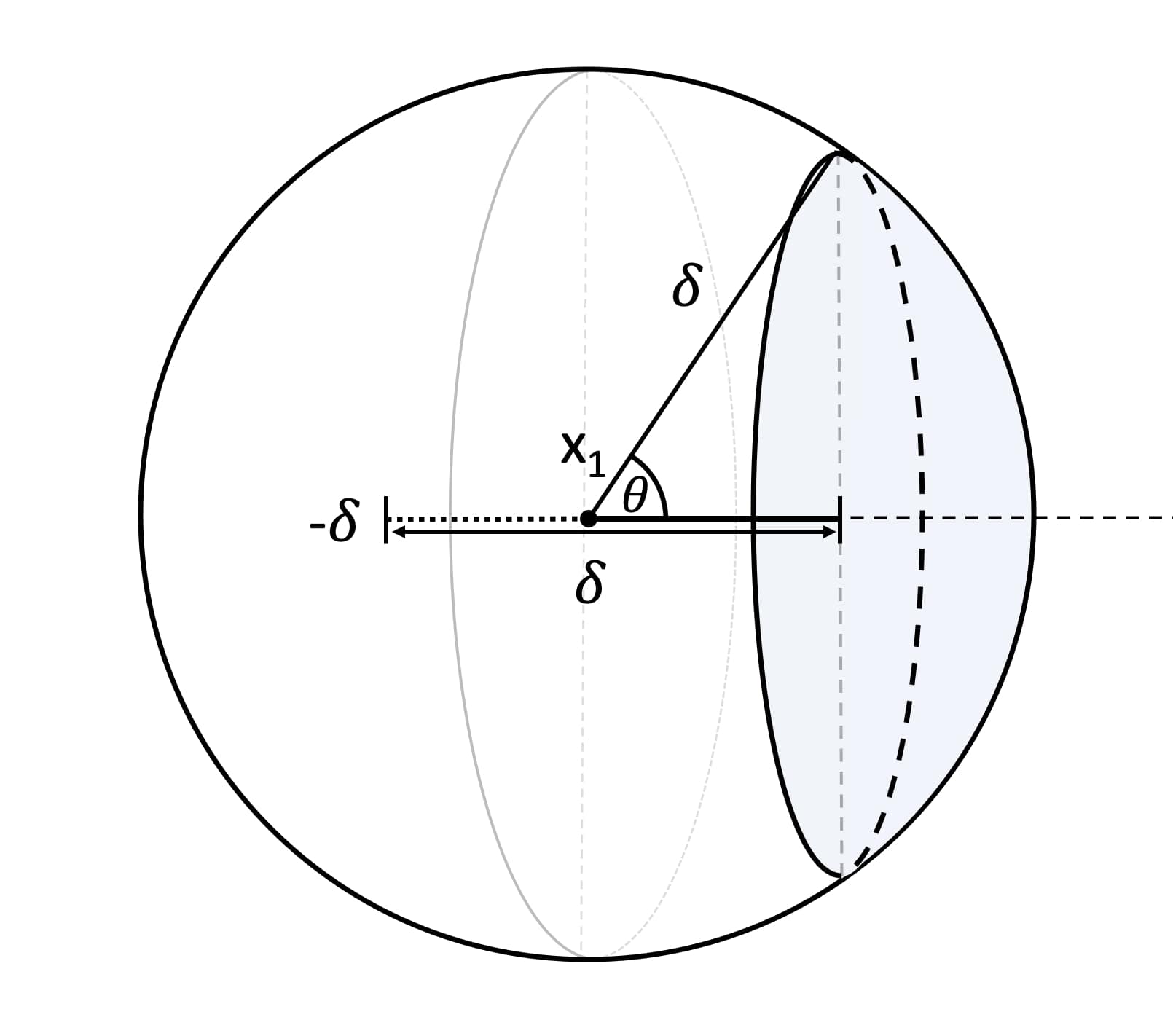}\par
\caption{An illustration of the parameters used for the probability of alpha escape function for 3-dimensions.}
\label{fig:3D}
\end{figure}

\begin{figure}[!t]
    \centering
    \includegraphics[width=\linewidth]{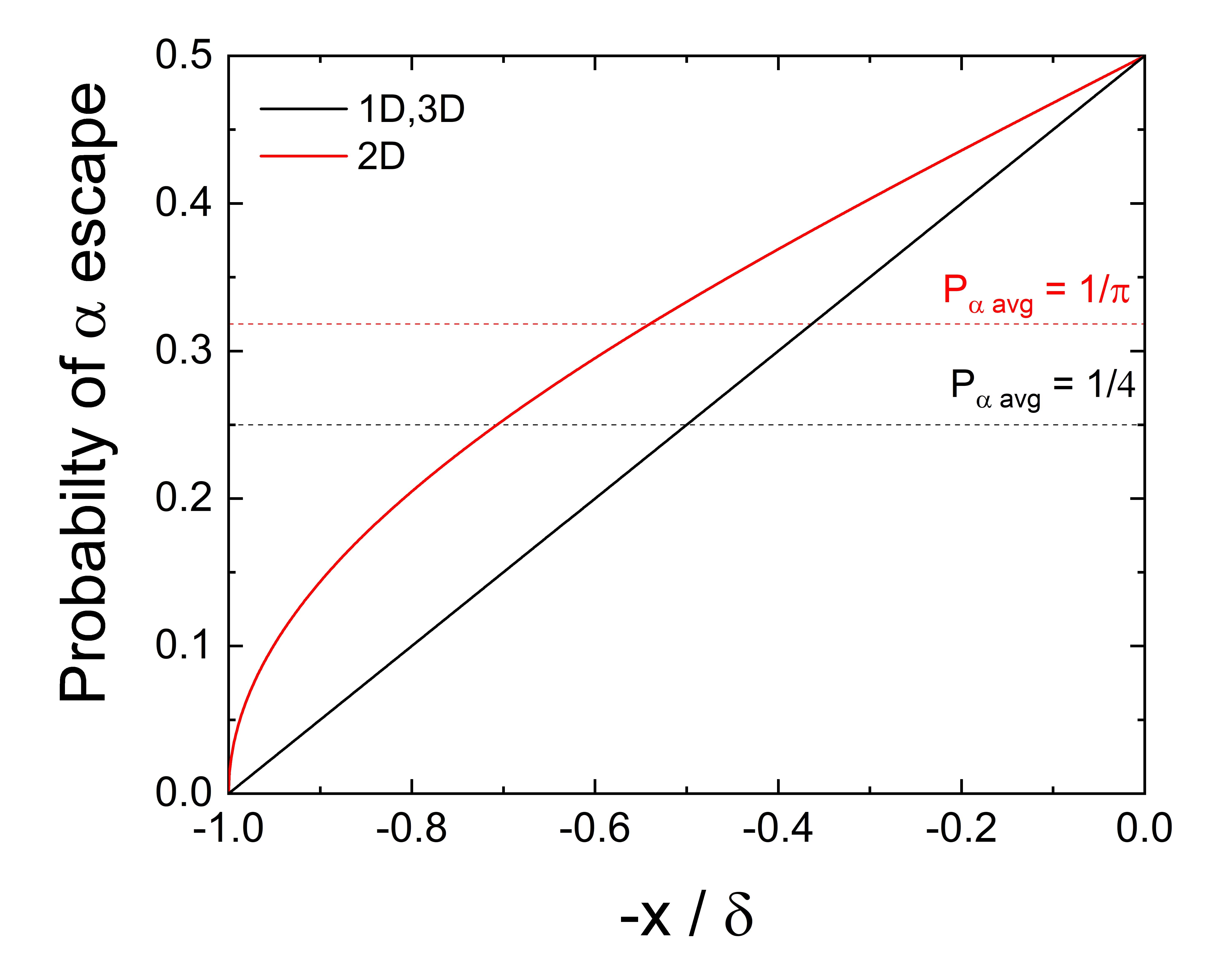}
    \caption{Probability of alpha particle escape as a function of fraction of maximal decay depth, $\delta$ in each of the three dimensions.}
    \label{fig:Dimen}
\end{figure}

Using the same reasoning for the derivation in 2-Dimensions we have a circle intersected by a line, as shown in Figure \ref{fig:2D}. The probability of the particle crossing the line becomes the length of the spherical arc past the line divided by all the possible end positions i.e the circles circumference. Using the formula
\begin{equation}
    Arclength = \pi \theta d
\end{equation}
And substituting for $x$ and $\delta$
\begin{equation}
    Arclength = 4 \pi \delta \arccos(\frac{|x|}{\delta}).
\end{equation}
Dividing through by the circumference of the circle, the probability of escape becomes
\begin{equation}
P = \frac{\arccos(|x|/\delta)}{\pi}.
\end{equation}
This is a non linear function, see Figure \ref{fig:Dimen}, that has an angular dependence and an average value of $1/\pi$. 

Lastly in a 3-Dimensional system the geometry becomes a sphere of radius, $\delta$, intersecting by a plane (Figure \ref{fig:3D}). Hence, the surface area of the cap beyond the plane, divided by the surface area of the sphere itself. Using the equation

\begin{equation}
    A_{cap} = 2\pi \delta h
\end{equation}
where h is the cap height. Substituting for $x$,
\begin{equation}
    A_{cap} = 2\pi \delta(\delta - |x|)
\end{equation}
then dividing by the surface are of the sphere, we arrive at the same probability function as the 1D model
\begin{equation}
P = \frac{1}{2} - \frac{|x|}{2\delta}.
\end{equation}
The probability of escape functions bound by the length $\delta$ are plotted in Figure \ref{fig:Dimen}

\vspace{10mm}

\newpage
\bibliographystyle{elsarticle-num}

\bibliography{main}

\begin{thebibliography}{10}
\expandafter\ifx\csname url\endcsname\relax
  \def\url#1{\texttt{#1}}\fi
\expandafter\ifx\csname urlprefix\endcsname\relax\def\urlprefix{URL }\fi
\expandafter\ifx\csname href\endcsname\relax
  \def\href#1#2{#2} \def\path#1{#1}\fi

\bibitem{innovation2020research}
NIRAB, Achieving net zero: The role of nuclear energy in decarbonisation,
  Nuclear Innovation and Research Advisory Board Annual Report (2020).

\bibitem{verbruggen2008renewable}
A.~Verbruggen, Renewable and nuclear power: A common future?, Energy Policy
  36~(11) (2008) 4036--4047.

\bibitem{seier2014environmental}
M.~Seier, T.~Zimmermann, Environmental impacts of decommissioning nuclear power
  plants: methodical challenges, case study, and implications, The
  International Journal of Life Cycle Assessment 19~(12) (2014) 1919--1932.

\bibitem{repository_2000}
L.~Johnson, P.~Smith, The interaction of radiolysis products and canister
  corrosion products and the implications for spent fuel dissolution and
  radionuclide transport in a repository for spent fuel, Tech. rep., National
  Cooperative for the Disposal of Radioactive Waste (NAGRA) (2000).

\bibitem{poinssot2005spent}
C.~Poinssot, C.~Ferry, M.~Kelm, J.~Cavedon, C.~Corbel, C.~Jegou, P.~Lovera,
  F.~Miserque, A.~Poulesquen, B.~Grambow, et~al., Spent fuel stability under
  repository conditions-final report of the european project, Tech. rep., CEA
  Saclay (2005).

\bibitem{shoesmith2007used}
D.~Shoesmith, Used fuel and uranium dioxide dissolution studies--a review,
  Nucl. Waste Manage. Org. Rep (2007).

\bibitem{bobrowski2017application}
K.~Bobrowski, K.~Skotnicki, T.~Szreder, Application of radiation chemistry to
  some selected technological issues related to the development of nuclear
  energy, in: Applications of Radiation Chemistry in the Fields of Industry,
  Biotechnology and Environment, Springer, 2017, pp. 147--194.

\bibitem{SHOESMITH1996287}
D.~Shoesmith, S.~Sunder, M.~Bailey, N.~Miller, Corrosion of used nuclear fuel
  in aqueous perchlorate and carbonate solutions, Journal of Nuclear Materials
  227~(3) (1996) 287 -- 299.

\bibitem{bright2019comparing}
E.~L. Bright, S.~Rennie, A.~Siberry, K.~Samani, K.~Clarke, D.~Goddard,
  R.~Springell, Comparing the corrosion of uranium nitride and uranium dioxide
  surfaces with h2o2, Journal of Nuclear Materials 518 (2019) 202--207.

\bibitem{SPRINGELL}
R.~Springell, S.~Rennie, L.~Costelle, J.~Darnbrough, C.~Stitt, E.~Cocklin,
  C.~Lucas, R.~Burrows, H.~Sims, D.~Wermeille, J.~Rawle, C.~Nicklin,
  W.~Nuttall, T.~Scott, G.~Lander, Water corrosion of spent nuclear fuel:
  radiolysis driven dissolution at the uo2/water interface, Faraday Discuss.
  180 (2015) 301--311.

\bibitem{buxton2008overview}
G.~V. Buxton, An overview of the radiation chemistry of liquids, Radiation
  chemistry: from basics to applications in material and life sciences EDP
  Sciences (2008).

\bibitem{Jonsson_steady}
F.~Nielsen, E.~Ekeroth, T.~E. Eriksen, M.~Jonsson, Simulation of radiation
  induced dissolution of spent nuclear fuel using the steady-state approach. a
  comparison to experimental data, Journal of nuclear materials 374~(1-2)
  (2008) 286--289.

\bibitem{Jonssonh2o2}
F.~Nielsen, K.~Lundahl, M.~Jonsson, Simulations of h2o2 concentration profiles
  in the water surrounding spent nuclear fuel, Journal of nuclear materials
  372~(1) (2008) 32--35.

\bibitem{Jonsson_time}
T.~E. Eriksen, M.~Jonsson, J.~Merino, Modelling of time resolved and long
  contact time dissolution studies of spent nuclear fuel in 10 mm carbonate
  solution--a comparison between two different models and experimental data,
  Journal of nuclear materials 375~(3) (2008) 331--339.

\bibitem{draganic1971radiation}
I.~Dragani{\'c}, Z.~Dragani{\'c}, E.~S. .~T. (Firm), The Radiation Chemistry of
  Water, Miami Winter Symposia, Academic Press, 1971.

\bibitem{elliot2009reaction}
A.~Elliot, D.~Bartels, The reaction set, rate constants and g-values for the
  simulation of the radiolysis of light water over the range 20 deg to 350 deg
  c based on information available in 2008, Tech. rep., Atomic Energy of Canada
  Limited (2009).

\bibitem{Nazhen_Shoesmith_Review}
N.~Liu, Z.~Zhu, L.~Wu, Z.~Qin, J.~J. No{\"e}l, D.~W. Shoesmith, Predicting
  radionuclide release rates from spent nuclear fuel inside a failed waste
  disposal container using a finite element model, Corrosion 75~(3) (2019)
  302--308.

\bibitem{grimes2017approximate}
D.~R. Grimes, D.~R. Warren, M.~Partridge, An approximate analytical solution of
  the bethe equation for charged particles in the radiotherapeutic energy
  range, Scientific reports 7~(1) (2017) 1--12.

\bibitem{bloch1933bremsvermogen}
F.~Bloch, Braking power {\ "o} genes of atoms with several electrons, magazine
  for {\ "u} r physics 81~(5) (1933) 363--376.

\bibitem{lindhard1963range}
J.~Lindhard, M.~Scharff, H.~E. Schi{\o}tt, Range concepts and heavy ion ranges,
  Munksgaard Copenhagen, 1963.

\bibitem{ziegler2008srim}
J.~Ziegler, J.~Biersack, M.~Ziegler, SRIM, the Stopping and Range of Ions in
  Matter, SRIM Company, 2008.

\bibitem{H_ion_Sore}
H.~S\o{}rensen, H.~H. Andersen, Stopping power of al, cu, ag, au, pb, and u for
  5---18-mev protons and deuterons, Phys. Rev. B 8 (1973) 1854--1863.

\bibitem{He_andersen}
H.~H. Andersen, H.~SøSrensen, The energy dependence of proton, deuteron, and
  helium-ion radiation damage in silver, platinum, and gold, Radiation Effects
  14~(1-2) (1972) 49--66.

\bibitem{H_ion_Au_ISHI}
R.~Ishiwari, N.~Shiomi, N.~Sakamoto, Stopping power of au for protons from 3 to
  8 mev, Nuclear Instruments and Methods in Physics Research Section B: Beam
  Interactions with Materials and Atoms 2~(1-3) (1984) 141--144.

\bibitem{montanari2017iaea}
C.~C. Montanari, P.~Dimitriou, The iaea stopping power database, following the
  trends in stopping power of ions in matter, Nuclear Instruments and Methods
  in Physics Research Section B: Beam Interactions with Materials and Atoms 408
  (2017) 50--55.

\bibitem{sunder1998calculation}
S.~Sunder, Calculation of radiation dose rates in a water layer in contact with
  used candu uo2 fuel, Nuclear technology 122~(2) (1998) 211--221.

\bibitem{nielsen2006geometrical}
F.~Nielsen, M.~Jonsson, Geometrical $\alpha$-and $\beta$-dose distributions and
  production rates of radiolysis products in water in contact with spent
  nuclear fuel, Journal of nuclear materials 359~(1-2) (2006) 1--7.

\bibitem{poulesquen2006spherical}
A.~Poulesquen, C.~Jegou, S.~Peuget, Determination of alpha dose rate profile at
  the uo 2/water interface, MRS Online Proceedings Library Archive 932 (2006).

\bibitem{hansson2020alpha}
N.~Hansson, C.~Ekberg, K.~Spahiu, Alpha dose rate calculations for uo2 based
  materials using stopping power models, Nuclear Materials and Energy 22 (2020)
  100734.

\bibitem{kumazaki2007determination}
Y.~Kumazaki, T.~Akagi, T.~Yanou, D.~Suga, Y.~Hishikawa, T.~Teshima,
  Determination of the mean excitation energy of water from proton beam ranges,
  Radiation Measurements 42~(10) (2007) 1683--1691.

\bibitem{miller2006MCNP}
W.~H. Miller, Dosimetry modeling for predicting radiolytic production at the
  spent fuel-water interface, Tech. rep., Yucca Mountain Project, Las Vegas,
  Nevada (2006).

\bibitem{mougnaud2015glass}
S.~Mougnaud, M.~Tribet, S.~Rolland, J.-P. Renault, C.~J{\'e}gou, Determination
  of alpha dose rate profile at the hlw nuclear glass/water interface, Journal
  of Nuclear Materials 462 (2015) 258--267.

\bibitem{tribet2017spent}
M.~Tribet, S.~Mougnaud, C.~J{\'e}gou, Spent nuclear fuel/water interface
  behavior: Alpha dose rate profile determination for model surfaces and
  microcracks by using monte-carlo methods, Journal of Nuclear Materials 488
  (2017) 245--251.

\bibitem{poulesquen2007linear}
A.~Poulesquen, C.~Jegou, Influence of alpha radiolysis of water on uo2 matrix
  alteration: chemical/transport model, Nuclear technology 160~(3) (2007)
  337--345.

\bibitem{dzaugis2015quantitative}
M.~E. Dzaugis, A.~J. Spivack, S.~D'Hondt, A quantitative model of water
  radiolysis and chemical production rates near radionuclide-containing solids,
  Radiation Physics and Chemistry 115 (2015) 127--134.

\bibitem{Cachior}
C.~Cachoir, P.~Carbol, J.~Cobos-Sabate, J.~Glatz, B.~Grambow, K.~Lemmens,
  A.~Martinez-Esparza, T.~Mennecart, C.~Ronchi, V.~Rondinella, et~al., Effect
  of alpha irradiation field on long-term corrosion rates of spent fuel, Spent
  fuel stability under repository conditions (2005).

\bibitem{hosoe1984stopping}
M.~Hosoe, Y.~Takami, F.~Shiraishi, F.~Tomura, Stopping power measurement using
  thick alpha sources, Nuclear Instruments and Methods in Physics Research
  223~(2-3) (1984) 377--381.

\bibitem{garisto1989energy}
F.~Garisto, The energy spectrum of $\alpha$-particles emitted from used
  candu™ fuel, Annals of Nuclear Energy 16~(1) (1989) 33--38.

\bibitem{JannsonJonsson}
M.~Jansson, M.~Jonsson, T.~Eriksen, Basic model of geometrical dose
  distributions from small uo2-particles, SKB-U-96-44 (1994) 6.

\end{thebibliography}

\end{document}